\documentclass[pre,floatfix,twocolumn,showpacs]{revtex4}
\usepackage{psfig}

\psfigurepath{/z/rue/couette:/z/rue/plots/schultzplot:
/net/dynamo/turu/natali/papers/cou2}
\renewcommand{\vec}[1]{\mbox{\boldmath $#1$}}
\def\Om{\Omega}
\def\q{\qquad}
\def\beg{\begin{eqnarray}}
\def\ende{\end{eqnarray}}
\def\gsim{\lower.4ex\hbox{$\;\buildrel >\over{\scriptstyle\sim}\;$}} 
\def\lsim{\lower.4ex\hbox{$\;\buildrel <\over{\scriptstyle\sim}\;$}}

\renewcommand{\textrm} [1] {\rm #1} 
\renewcommand{\textit} [1] {\it #1}
\renewcommand{\textbf} [1] {\bf #1} 

\begin{document}

\title{The  linear MHD Taylor-Couette instability for liquid sodium}
\author{G\"unther R\"udiger,  Manfred Schultz, Dima Shalybkov\footnote{permanent address: A.F. Ioffe Institute for Physics and Technology, 194021, St.
Petersburg, Russia}}
\affiliation{Astrophysikalisches Institut Potsdam,
         An der Sternwarte 16, D-14482 Potsdam, Germany}
\email{gruediger@aip.de}
\date{\today}

\begin{abstract}
The linear stability of MHD Taylor-Couette flow of infinite
vertical extension   is considered for liquid sodium with its small magnetic Prandtl number Pm of order of $10^{-5}$.
The calculations are  performed for a  container with 
$\hat\eta=0.5$ with an axial uniform magnetic field excluding counter rotating 
cylinders. The sign of the constant $a$ in the basic rotation law  $\Omega=a+b/R^2$ strongly influences the presented results. It is negative for  resting outer cylinder. The main point here is that   the subcritical excitation which occurs for large Pm disappears for 
small Pm (cf. Fig. \ref{minima}).  This is the reason that the existence of the magnetorotational instability remained unknown over decades.

For rotating outer cylinder the limiting case $a=0$ (i.e. $\hat \mu = \hat \eta^2$) plays an exceptional role. The hydrodynamic instability starts to disappear while the hydromagnetic instability exists with minimal Reynolds numbers  at certain Hartmann numbers of the magnetic field. These  Reynolds numbers exactly scale  with Pm$^{-1/2}$ resulting in moderate values of order $10^4$ for Pm=10$^{-5}$.  However, already for the smallest positive value  of $a$ the Reynolds numbers start to scale as 1/Pm  leading  to  much higher values of order $10^6$ for Pm=10$^{-5}$. Hence, for outer cylinders rotating faster than the limit $a=0$ it is exclusively  the {\em magnetic}  Reynolds number Rm  which directs the excitation of the instability. They are resulting as lower for 
  insulating walls (`vacuum')
 than for conducting walls.  Generally, the magnetic Reynolds 
numbers for liquid sodium have to exceed values of order 10 leading  to frequencies of about   20 Hz for the 
rotation of the inner cylinder if containers with  (say) 10 cm   radius  are considered. The required magnetic fields are  about 1000 Gauss.

Also  nonaxisymmetric  modes  have  been 
considered.  With vacuum boundary conditions their excitation is always 
more difficult than the excitation of   axisymmetric modes; we never 
observed a crossover of the  lines of marginal stability. For conducting walls, 
however,  such crossovers exist  for both resting and rotating outer 
cylinders,  and this  might be essential for future dynamo experiments. In this case, however, the instability  also can onset in form of {\em oscillating} axisymmetric patterns of flow and field and the Reynolds numbers of these solutions are lower than the Reynolds numbers for the stationary solutions.

\end{abstract}


\maketitle
\section{Introduction}
The longstanding problem of the generation of turbulence in various
hydrodynamically stable situations has found a solution in recent years
with the MHD shear flow instability, also called magnetorotational instability (MRI), in
which the presence of a magnetic field has a destabilizing effect on a
differentially rotating flow with the angular velocity decreasing
outwards.
The MRI has been  formulated decades ago \cite{V59,C61} for
ideal Taylor-Couette flow, but its importance as the source of
turbulence in accretion discs with differential (Keplerian) rotation
was first recognized by Balbus and Hawley, \cite{BH91}.

However, the MRI has never been observed in the laboratory
\cite{DO60,DO62,DC64,B70}.  
After Goodman and Ji \cite{GJ01} was the absence of MRI
 due to the small magnetic Prandtl number approximation used in
\cite{C61}. 
The magnetic Prandtl number Pm is really very small under laboratory conditions
($\sim 10^{-5}$ and smaller, see Table \ref{tab0}).  

\begin{table}
\caption{\label{tab0} Parameters of the fluids suitable for MHD
experiments taken from \cite{C61} and \cite{NPCNB02}}
\medskip
\begin{tabular}{|l|c|c|c|c|}
\hline
 & $\rho$ [g/cm$^3$] & $\nu$ [cm$^2$/s] & $\eta$ [cm$^2$/s] & Pm\\[0.5ex]
\hline
Mercury & 5.4 & 1.1$\cdot 10^{-3}$ & 7600 & 1.4$\cdot
10^{-7}$\\[0.5ex]
\hline 
 Gallium & 6.0 & 3.2$\cdot 10^{-3}$ & 2060 & 1.5$\cdot 10^{-6}$\\[0.5ex]
 \hline
 Sodium & 0.92 & 7.1$\cdot 10^{-3}$ & 810 & 0.88$\cdot 10^{-5}$\\[0.5ex]
 \hline
 \end{tabular}
 \end{table}
A proper understanding of this phenomenon
is very important for possible future experiments, including
Taylor-Couette flow  dynamo experiments.
\begin{figure}
\psfig{figure=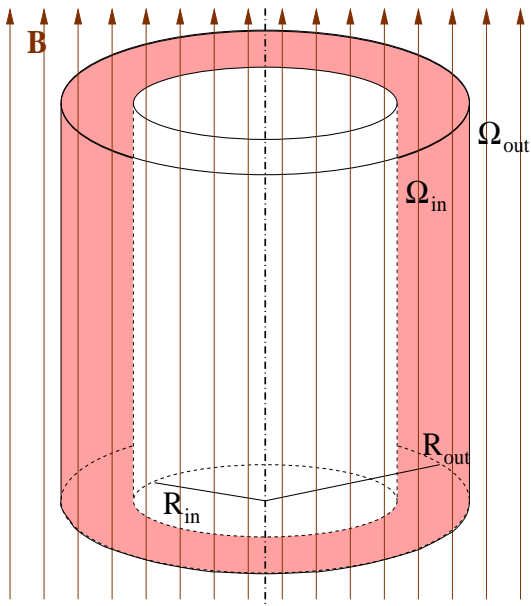,width=7cm,height=8cm}
\caption{\label{geometry} Cylinder geometry of the  Taylor-Couette flow
with axial magnetic field.}
\end{figure}
The simple model of uniform  density fluid
contained between two vertically-infinite rotating cylinders is used with 
constant magnetic field parallel to the rotation axis. 
For viscous flows the most general form of the rotation law $\Om(R)$ in the
fluid is
\begin{equation}
{\Om}(R) = a+b/{R}^2,
\label{Om}
\end{equation}
where $a$ and $b$ are two constants related to the angular
velocities $\Om_{\textrm{in}}$ and $\Om_{\textrm{out}}$ with which the inner
and the outer cylinders are rotating and $R$ is the distance from the
rotation axis. If $R_{\textrm{in}}$ and $R_{\textrm{out}}$
($R_{\textrm{out}}>R_{\textrm{in}}$) are the radii of the two cylinders then
\begin{equation}
a={\hat \mu-{\hat\eta}^2\over1-{\hat\eta}^2}{\Om}_{\rm in}
\ \ \ \ \ {\textrm{and}} \  \ \ \  
b= R_{\textrm{in}}^2 {1-\hat\mu \over1-{\hat\eta}^2} {\Om}_{\rm in}
\label{ab}
\end{equation}
with 
\begin{equation}
\hat\mu={\Om}_{\textrm{out}}/\Om_{\textrm{in}}  \q  {\textrm{and}} \q 
\hat\eta=R_{\textrm{in}}/R_{\textrm{out}}.
\label{mu}
\end{equation}
Following the Rayleigh stability criterion, $d(R^2 {\Om})^2/dR>0$,
rotation laws are hydrodynamically stable
for $a>0$, i.e. $\hat\mu>\hat\eta^2$. They should in particular  be stable for
resting inner cylinder, i.e. $\hat\mu \to \infty$. Richard and Zahn 
\cite{RZ99} focused
attention to the experimental results of Wendt \cite{W1933} who found {\em nonlinear} 
instability for this case for Reynolds numbers of order 10$^5$. The
finite-amplitude instability of hydrodynamically stable rotation laws must
therefore
remain in the astrophysical discussion. However, later experiments with very similar
Taylor-Couette flow experiments for resting inner cylinder demonstrated  the results
of Wendt as due to rather imperfect container constructions and the flow
remained laminar even for Reynolds numbers up to 10$^6$,  \cite{Sch59}.

\begin{figure}
\psfig{figure=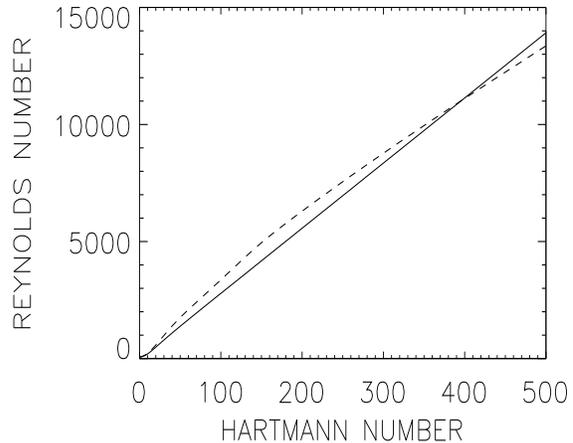,width=8cm,height=7cm}
\caption{\label{old} Instability lines for axisymmetric (solid) and nonaxisymmetric  modes (m=1,  dashed line) for conducting walls and Pm=0.01 ($\hat\eta=0.5$).}
\end{figure}

One of the targets in the present paper is the axisymmetry of the excited
modes. We have shown in \cite{SRS02}  that for containers with conducting
boundaries it happens for sufficiently strong magnetic fields that the mode
with the lowest eigenvalue (i.e. the lowest Reynolds number) is a
nonaxisymmetric mode. As an impressive example,  in Fig. \ref{old} for 
Pm=0.01 the crossover of the instability lines for axisymmetric and 
nonaxisymmetric modes is shown for Hartmann numbers of about 400 
(see \footnote{Note that 
we considered in \cite{SRS02}  only  the excitation of the instability for 
Hartmann number smaller than about 100}). 
Despite of its general meaning this behavior 
 is only known
so far for conducting walls and for magnetic Prandtl numbers not smaller
than 10$^{-2}$ (see \cite{KGDIP66}). For possible laboratory experiments we have to extend, however,  the
computations to insulating boundaries (vacuum) and to much smaller magnetic
Prandtl numbers Pm.

The equations, therefore, are 
solved here mainly  for the single small magnetic Prandtl number 
Pm $=10^{-5}$ very close to the value for liquid sodium (see Table
\ref{tab0}). The
aspect ratio of the container walls radii in the present paper is fixed to $\hat\eta
= 0.5$.
\section{Basic equations}
The MHD equations which have to be solved are 
\beg
{\partial \vec{u} \over \partial t} + (\vec{u} \nabla)\vec{u} = - {1\over \rho}
\nabla p + \nu \Delta \vec{u} + \vec{J} \times \vec{B}
\label{0}
\ende
and
\beg
{\partial \vec{B} \over \partial t}= {\rm curl} (\vec{u} \times \vec{B}) + \eta \Delta
\vec{B},
\label{0.1}
\ende
with the electric current $J= {\rm curl} \vec{B}/\mu_0$ and
$
{\rm div} \vec{u} = {\rm div} \vec{B} = 0$.
They are considered in cylindrical geometry with 
$R$, $\phi$, and $z$ as the cylindrical coordinates.  A viscous 
electrically-conducting incompressible fluid between
two rotating infinite cylinders in the presence of a uniform magnetic
field  parallel to the rotation axis leads to the basic solution
$U_R=U_z=B_R=B_\phi=0,
B_z=B_0={\rm const.,} \ {\rm and} \  U_\phi=a R+b/R$, 
with  $\vec{U}$ as  the flow  and  $\vec{B}$ as the magnetic field. 
We are interested in the stability of
this solution. The perturbed state of the flow may be  described by
$u_R', \; u_\phi', \; u_z', \; B_R', \; B_\phi', \; B_z',
\; p'$
with  $p'$ as the  pressure perturbation.

Here only  the linear stability problem is considered.
By analyzing the disturbances into normal modes  the solutions
of the linearized hydromagnetic equations are of the form
\begin{eqnarray}
 \vec{B}'&=& \vec{B}(R)\ {\rm e}^{{\rm i}(m\phi+kz-\omega t)},  \nonumber \\
\vec{u}'&=&\vec{u}(R)\ {\rm e}^{{\rm i}(m\phi+kz-\omega t)}. 
\end{eqnarray}
From hereon all dashes  have been omitted from the  notations of fluctuating quantities. Only marginal stability  will be considered hence  the imaginary part of 
 $\omega$, i.e. ${\cal I}(\omega)$, always 
 vanishes.
We use 
\begin{equation}
H=\sqrt{R_{\rm in}(R_{\rm out} - R_{\rm in})}
\label{2.4}
\end{equation}
 as the unit of length,
the  $ \eta /H$ as the unit of
perturbed velocity and $B_0 $ as the unit of perturbed
magnetic field and work with the magnetic Prandtl number 
\begin{equation}
{\rm Pm} = {\nu\over\eta},
\label{pm}
\end{equation}
 with $\nu$  the
kinematic viscosity and $\eta$  the magnetic diffusivity. 
Note  $H^{-1}$ also as the unit of wave numbers and $\nu/H^2$ as the unit of frequencies.
After elimination of both pressure fluctuations and the fluctuations of the vertical magnetic field, $B'_z$, the linearized equations are
\beg
{\partial u_R \over \partial R} + {u_R \over R} + {{\rm i}m \over R} u_\phi + 
{\rm i}k u_z = 0,
\label{1}
\ende
\begin{eqnarray}
\lefteqn{{\partial^2 u_\phi \over \partial R^2} + {1\over R} {\partial u_\phi
\over \partial R} - {u_\phi \over R^2} - \left({m^2 \over R^2} + k^2\right)
u_\phi -}\nonumber\\
\lefteqn{-{\rm i} \left(m {\rm Re} {\Om \over \Om_{\rm in}} - \omega\right)
u_\phi + {2{\rm i}m \over R^2} u_R - {\rm Re} {1\over R} 
{\partial \over \partial R}
\left(R^2 {\Om \over \Om_{\rm in}}\right) u_R} \nonumber\\
&& - {m \over k} \left[{1\over R} {\partial^2 u_z \over \partial R^2} + 
{1\over R^2} {\partial u_z \over \partial R} - \left({m^2 \over R^2} + 
k^2\right) {u_z
\over R} - \right. \nonumber\\
&& \left. - {\rm i}\left(m {\rm Re} {\Om \over \Om_{\rm in}} - \omega\right) 
{u_z
\over R}\right]  + {m\over k} {\rm Ha}^2 \left[{1\over R} {\partial B_R \over 
\partial R} +
{B_R \over R^2}\right] +\nonumber\\
&& + {{\rm i}\over k} {\rm Ha}^2 \left({m^2\over R^2} + k^2\right)
B_\phi = 0,
\label{2}
\end{eqnarray}
\begin{eqnarray}
\lefteqn{{\partial^3 u_z \over \partial R^3} + {1\over R} 
{\partial^2 u_z \over \partial
R^2} - {1\over R^2} {\partial u_z \over \partial R} - \left({m^2\over R^2} +
k^2\right) {\partial u_z \over \partial R} +}\nonumber\\
\lefteqn{+{2m^2 \over R^3} u_z - {\rm i}\left(m
{\rm Re} {\Om\over \Om_ {\rm in}} - \omega\right) {\partial u_z \over
\partial R} -  
 {\rm i}m {\rm Re} {\partial \over \partial R} \left({\Om \over \Om_{\rm
in}}\right) u_z} \nonumber\\ 
&& - {\rm Ha}^2 \left[{\partial^2 B_R \over \partial R^2} + {1\over
R} {\partial B_R \over \partial R} - {B_R \over R^2} - k^2 B_R + \right. \nonumber\\
&& \left. +{{\rm i}m \over R}
{\partial B_\phi \over \partial R} - {{\rm i}m\over R^2} B_\phi \right] -
{\rm i}k\left[{\partial^2 u_R \over \partial R^2} + {1\over R} {\partial u_R
\over \partial R} - {u_R \over R^2} - \right. \nonumber\\
&& \left. - \left(k^2 + {m^2\over R^2}\right)
u_R\right] - k \left(m {\rm Re} {\Om \over \Om_{\rm in}} - \omega\right)
u_R -\nonumber\\
&& - 2 {km \over R^2} u_\phi - 2 {\rm i}k {\rm Re} {\Om \over \Om_{\rm in}}
u_\phi = 0,
\label{3}
\end{eqnarray}
\begin{eqnarray}
\lefteqn{{\partial^2 B_R \over \partial R^2} + {1\over R} {\partial B_R \over 
\partial R}
- {B_R \over R^2} - \left({m^2 \over R^2} + k^2\right) B_R -}\nonumber\\
&& - {2{\rm i}m\over R^2}
B_\phi - {\rm i} {\rm Pm} \left(m {\rm Re} {\Om \over \Om_{\rm in}}
-\omega\right) B_R + {\rm i}k u_R=0,
\label{4}
\end{eqnarray}
\begin{eqnarray}
\lefteqn{{\partial^2 B_\phi \over \partial R^2} + {1\over R} 
{\partial B_\phi \over
\partial R} -{B_\phi \over R^2} - \left({m^2 \over R^2} + k^2\right) B_\phi
+}\nonumber\\
&& +{2{\rm i}m \over R^2} B_R -{\rm i} {\rm Pm} \left(m {\rm Re} 
{\Om \over \Om_{\rm in}}
- \omega\right) B_\phi + {\rm i}k u_\phi+\nonumber\\ 
&& + {\rm Pm} \ {\rm Re} \ R {\partial \Om
/\Om_{\rm in} \over \partial R} B_R = 0.
\label{5}
\end{eqnarray}
Here the Reynolds number Re and the Hartmann number Ha are defined as
\begin{equation}
 {\rm Re} = {{\Om}_{\rm
in} R_{\rm in} (R_{\rm out} - R_{\rm in}) \over \nu}
\label{RE}
\end{equation}
and
\begin{equation}
{\rm Ha} = 
B_0 \sqrt{R_{\rm in}(R_{\rm out} - R_{\rm in}) \over 
\mu_0 \rho \nu \eta}.
\label{HA}
\end{equation}
For  given Hartmann number and magnetic Prandtl number in the present paper we 
shall compute with a linear theory the critical Reynolds number of the 
rotation of the inner cylinder, also for various mode numbers $m$.
\section{Boundary conditions, numerics}
An appropriate set of ten boundary conditions is needed to solve  the system
(\ref{1})$\dots$(\ref{5}). 
Always no-slip conditions for the velocity on the walls
are used, i.e. 
$
u_R=u_\phi=d u_R/dR=0.$
The boundary conditions for the magnetic field depend on the electrical properties
of the walls. The tangential currents and the radial component of the 
magnetic field  vanish on conducting walls  hence
$
d B_\phi/dR + B_\phi/R = B_R = 0.$
These   boundary conditions  may also hold  both 
for $R=R_{\rm in}$ and  for $R=R_{\rm out}$.

The homogeneous set of equations (\ref{1})$\dots$(\ref{5}) together with the boundary
conditions determine the eigenvalue problem of the form 
$
{\cal L}(k, m,  {\rm Re}, {\rm Ha}, {\cal {R}}(\omega))=0
$ 
for given Pm. The real part of $\omega$, i.e.  ${\cal {R}}(\omega)$, describes a  
drift of the pattern along the azimuth which only exists for nonaxisymmetric 
flows. For axisymmetric flows $(m=0)$ the real part of $\omega$, i.e.
${\cal R}(\omega)$, is zero for stationary patterns of flow and field and it 
is nonzero for oscillating solutions, which are called overstability. $\cal L$ is a complex quantity, both its real part and its imaginary part 
must vanish for the critical Reynolds number. The latter is 
minimized by choice of the wave number $k$. ${\cal R}(\omega)$ is the second
quantity which is fixed by the eigen equation.

The system is approximated by finite differences with typically 200 radial grid points. The
resulting determinant, $\cal L$, takes the value zero if and only if the 
values  Re  are the eigenvalues. We can also stress that the results are numerically  robust as an  increase of the number of grid points does not change the results remarkably.  
For a fixed Hartmann number, a fixed Prandtl number and a given vertical wave 
number $k$ we find the eigenvalues of the equation system. They are always minimal 
for a certain wave number which by itself defines the marginally unstable mode.
The corresponding eigenvalue is the desired Reynolds number.

The situation changes for insulating  walls.
The magnetic field must  match the external magnetic field for
vacuum. It is known for this case that the boundary conditions for axisymmetric
solutions strongly differ from those for nonaxisymmetric solutions (see
 \cite{EMR90}). The condition ${\rm curl}_R{\vec B}=0$ in vacuum immediately
provides  
\beg
B_\phi={m\over kR} B_z
\label{bphi}
\ende
at $R=R_{\textrm{in}}$ and $R=R_{\rm out}$. From the solution of the
potential equation $\Delta \psi=0$ one finds 
\beg
B_R+ {{\rm i} B_z \over I_m(kR)} \left({m\over kR} I_m(kR)+I_{m+1}(kR)\right)=0
\label{BR1}
\ende
for $R=R_{\textrm{in}}$ and  
\beg
B_R+{{\rm i} B_z \over K_m(kR)} \left({m\over kR} K_m(kR) - K_{m+1}(kR)\right)=0
\label{B_R2}
\ende
for $R=R_{\rm out}$. $I_m$ and $K_m$ are the modified Bessel functions
(with different behavior at $R \to 0$ and $R\to \infty$).  One can  eliminate with  div${\vec B}$=0  the vertical component $B_z$ of the magnetic field in the boundary conditions (\ref{bphi})...(\ref{B_R2}).


\section{Results }
The following results concern different aspects of the MHD Taylor-Couette
problem for small magnetic Prandtl number Pm. In Section A the main
realization of the case $a<0$ (here with resting outer cylinder, i.e.
$\hat\mu =0$) is considered. There is instability even without magnetic
fields so that the bifurcation lines start at the y-axis. In Section B the
special case $a=0$ is considered with very surprising results. The Section
C presents the results for the two experiments with $\hat\mu=0$ and
$\hat\mu=0.33$ with respect to the axisymmetry of the eigenmodes. In
Section D the existence of oscillating modes is discussed, i.e. the case of
overstability for small magnetic Prandtl numbers.
\subsection{Subcritical excitation for large  Pm ($\vec{a}{\bf <0}$)}
\begin{figure}
\psfig{figure=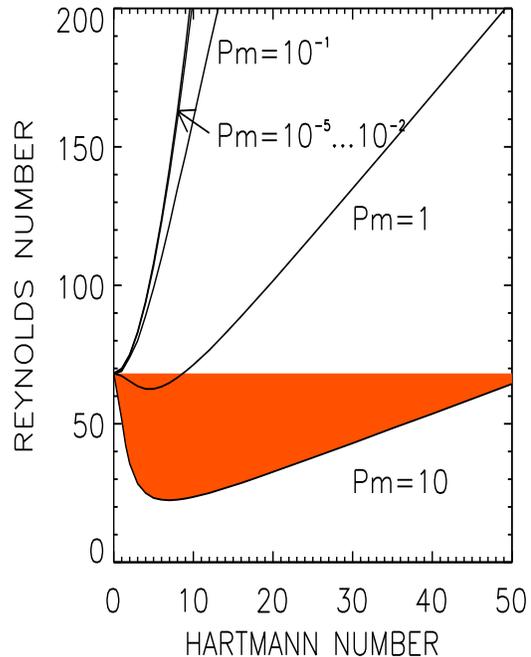,width=8cm,height=10cm}
\caption{\label{minima} Marginal stability lines for axisymmetric modes 
with resting outer cylinder of conducting material. The shaded area denotes
subcritical excitations of unstable axisymmetric modes by the external
magnetic field.}
\end{figure}
Figure \ref{minima} shows the stability lines for axisymmetric modes for
containers with conducting walls and with resting outer cylinder for 
fluids of various magnetic
Prandtl number. Only the vicinity of the classical hydrodynamic solution
with Re $= 68$ is shown. There is a strong difference of the geometry of
the bifurcation lines for Pm $\gsim 1$ and Pm $<1$. In the latter case, for
fluids with low electrical conductivity the magnetic field only
suppresses the instability so that all the critical Reynolds numbers exceed
the value 68, and this the more the stronger the magnetic field is.

For sufficiently small magnetic Prandtl number the stability lines hardly
differ, which is the situation already considered by Chandrasekhar 
\cite{C61} without any indication of magnetorotational instability.

The opposite is true for Pm $\gsim 1$. Note that in Fig. \ref{minima} for
materials with high electrical conductivity the resulting critical Reynolds
numbers are smaller than Re $=68$. The magnetic field with small Hartmann
numbers support instability patterns rather than to suppress them. This
effect becomes more effective for increasing Pm but it vanishes for
stronger magnetic fields. Obviously, the 
MRI only exists for weak magnetic fields and high enough electrical conductivity and/or molecular viscosity 
(when the fields can be considered as frozen in and/or enough  viscosity 
prevents the action of the Taylor-Proudman theorem).

\begin{figure}[ht]
\psfig{figure=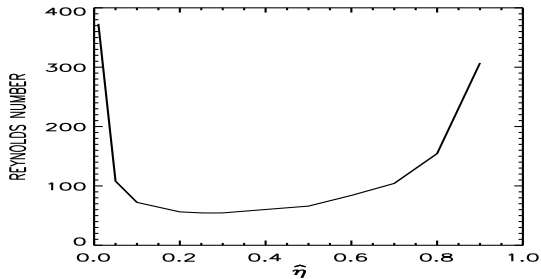,width=8cm,height=4cm}
\caption{\label{reeta} Critical Reynolds number versus $\hat\eta$
for $\hat\mu=\hat\eta^2$ and Pm$=1$.}
\label{fig4}
\end{figure}

Note that the subcritical excitation of Taylor vortices only works for weak
magnetic fields. The upper limits of the possible Hartmann numbers can be
observed for the magnetic Prandtl numbers 1 and 10 in Fig. \ref{reeta}.
After our computations the subcritical excitation of Taylor vortices for
weak magnetic fields requires rather high magnetic Prandtl numbers. The
microscopic values for Pm are orders of magnitudes smaller than unity, so
that there should be no chance to realize the subcritical excitation of
Taylor vortices by experiments. However, the speculation may be allowed
whether really the microscopic Pm is the basic input. The scenario is also
interesting whether possible finite-amplitude hydrodynamic instabilities
provide some kind of background turbulence which can be considered as
modifying  the value of the magnetic Prandtl number, \cite{NPCNB02}. The
turbulence influences both the viscosity values and the magnetic-diffusivity
values so that
\beg
{\textrm Pm} \to {\textrm Pm}_{\textrm eff} = {\nu + \nu_{\textrm T} \over
\eta + \eta_{\textrm T}} \simeq {\nu_{\textrm T} \over \eta_{\textrm T}}
\label{Pmeff}
\ende
with $\nu_{\textrm T}$ and $\eta_{\textrm T}$ as the eddy viscosity and the
eddy diffusivity, resp. Because of the existence of the pressure term in
the momentum equation, both quantities are not identical. We do not have
precise knowledge about the effective turbulent magnetic Prandtl number but
 it has been demonstrated that values of order 0.1 or somewhat larger should
 not be unlikely, \cite{R89}. Insofar if such speculations are not too far
 from the reality, it is not completely clear that the subcritical
 excitation of Taylor vortices which we have presented in Fig. \ref{reeta} is
 unobservable in general.  

\subsection{The case ${\bf a=0}$ ($\vec{\hat \mu} = \vec{\hat \eta^2}$)}
There is a universal scaling on Pm for the special case with $a=0$ in the
basic flow profile (\ref{Om}), i.e. for $\hat\mu =\hat\eta^2$. Then the term
with $\partial(R^2\Om)/\partial R$ in Eq. (\ref{2}) vanishes and for
$m=\omega=0$ one finds that the quantities $u_R, u_z, B_R$ and $B_z$ are 
scaling
as Pm$^{-1/2}$ while $u_\phi, B_\phi, k$   Ha scale as Pm$^0$. Then also the Reynolds number for the axisymmetric modes scales as 
\beg
{\rm Re} \propto {\rm Pm}^{-1/2}.
\label{Re}
\ende
The scaling does not depend on the boundary
conditions as these for $m=0$ also comply with  the relations.

The result (\ref{Re}) has numerically been found by Willis and Barenghi 
 for vacuum boundary conditions, \cite{WB02}. However, R\"udiger and 
 Shalybkov \cite{RS02} for $a > 0$ ($\hat \mu > \hat\eta^2$) found the much steeper
scaling
\beg
{\rm Re} \propto {\rm Pm}^{-1},
\label{RePm}
\ende
resulting in the surprisingly simple relation
\beg
{\rm Rm} \propto {\rm const.}
\label{REPM}
\ende
for the magnetic Reynolds number ${\rm Rm} = {{\Om}_{\rm
in} R_{\rm in} (R_{\rm out} - R_{\rm in})/ \eta}$ and
\beg
{\rm Ha} \propto {\rm Pm}^{-1/2}
\label{Hpro}
\ende
resulting in
\beg
{\rm Ha}^* \propto {\rm const.}
\label{Hapro}
\ende
for 
${\rm Ha}^*=B_0\sqrt{R_{\rm in}(R_{\rm out} - R_{\rm in})/\mu_0 \rho \eta^2}
$ 
(Lundquist number,   see \cite{RS02}). In case of small magnetic Prandtl number the exact value of 
 the microscopic viscosity is totally unimportant for the excitation of 
 the instability. In consequence, however, the corresponding Reynolds numbers for the MRI seem to differ by 2 orders of magnitude, i.e.  
10$^4$  and 10$^6$. Insofar, experiments with $\hat\mu =\hat\eta^2$ seem 
to look much more promising than experiments with $\hat \mu > \hat\eta^2$. 

\begin{figure}[ht]
\hbox{
\psfig{figure=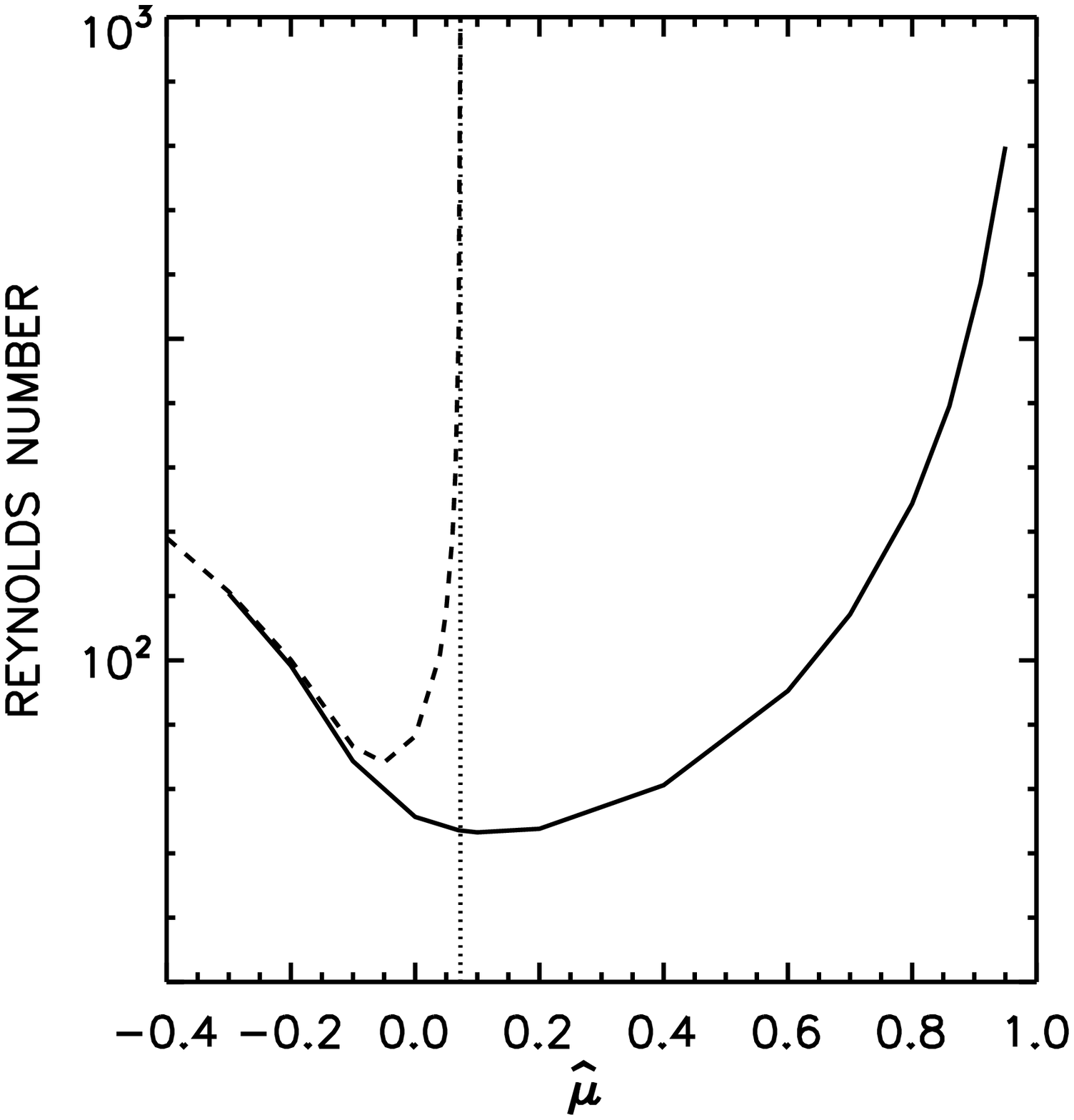,width=4.4cm,height=8cm}\hfill
\psfig{figure=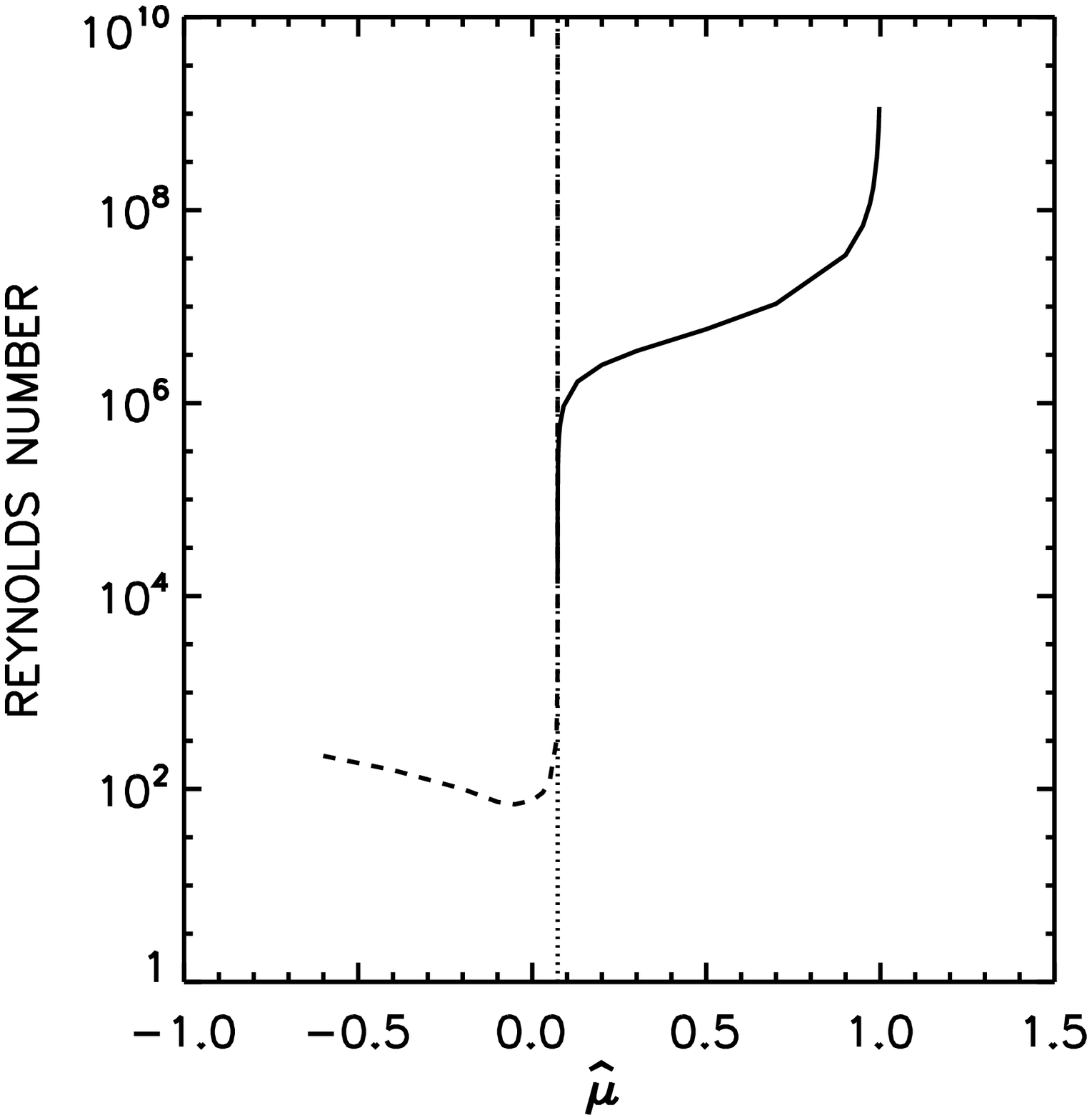,width=4.4cm,height=8cm}}
\caption{\label{remu} Critical Reynolds numbers for the Taylor-Couette flow versus $\hat\mu$
for $\hat\eta=0.27$ and  Pm=1 (left) and  Pm=$10^{-5}$ (right).
The  curve for the hydrodynamic  instability (Ha=0) is dashed and  the 
hydromagnetic   curve (Ha$>$0)  is solid.  The  dotted line denotes the location 
of  $a$=0, i.e. $\hat\mu=\hat\eta^2$.  }
\label{fig5}
\end{figure}

Unfortunately,   this   challenging possibility cannot be utilized  in experiments.
The critical Reynolds number for $\hat\mu = \hat\eta^2$ and Pm=1 as a
function of $\hat\eta$ is given in Fig. \ref{fig4}. The total minimum of the Reynolds number is  54.4 
for $\hat\eta = 0.27$ so that after (\ref{Re})  one 
expects the value 1.7$\cdot 10^4$ for the Reynolds number for Pm$=10^{-5}$. Fig. \ref{fig5} shows
the behavior of this result in the vicinity of $\hat\mu = \hat\eta^2$.
There is a vertical jump from 10$^4$ to 10$^6$ in an extremely small
interval of the abscissa. This sharp transition does not exist for Pm=1, 
it is only due to the very small
value of Pm.  For this case in Fig. \ref{fig5}  the
coexistence of both hydrodynamic and hydromagnetic instability is also 
presented.
The jump profile for Pm$=10^{-5}$ in Fig. \ref{fig5} (right) makes it clear that such 
experiments with $\hat\mu = \hat\eta^2$ are not possible. Even the smallest
deviation from the condition $\hat\mu = \hat\eta^2$ drastically changes the
excitation condition. For $\hat\mu$ smaller than $\hat\eta^2$ (negative
deviations) the hydrodynamic instability sets in and for $\hat\mu$ slightly 
exceeding $\hat\eta^2$ (positive deviations) the Reynolds number suddenly
jumps by two orders of magnitudes.

\subsection{Excitation of nonaxisymmetric modes}
\begin{figure}[ht]
\hbox{
\psfig{figure=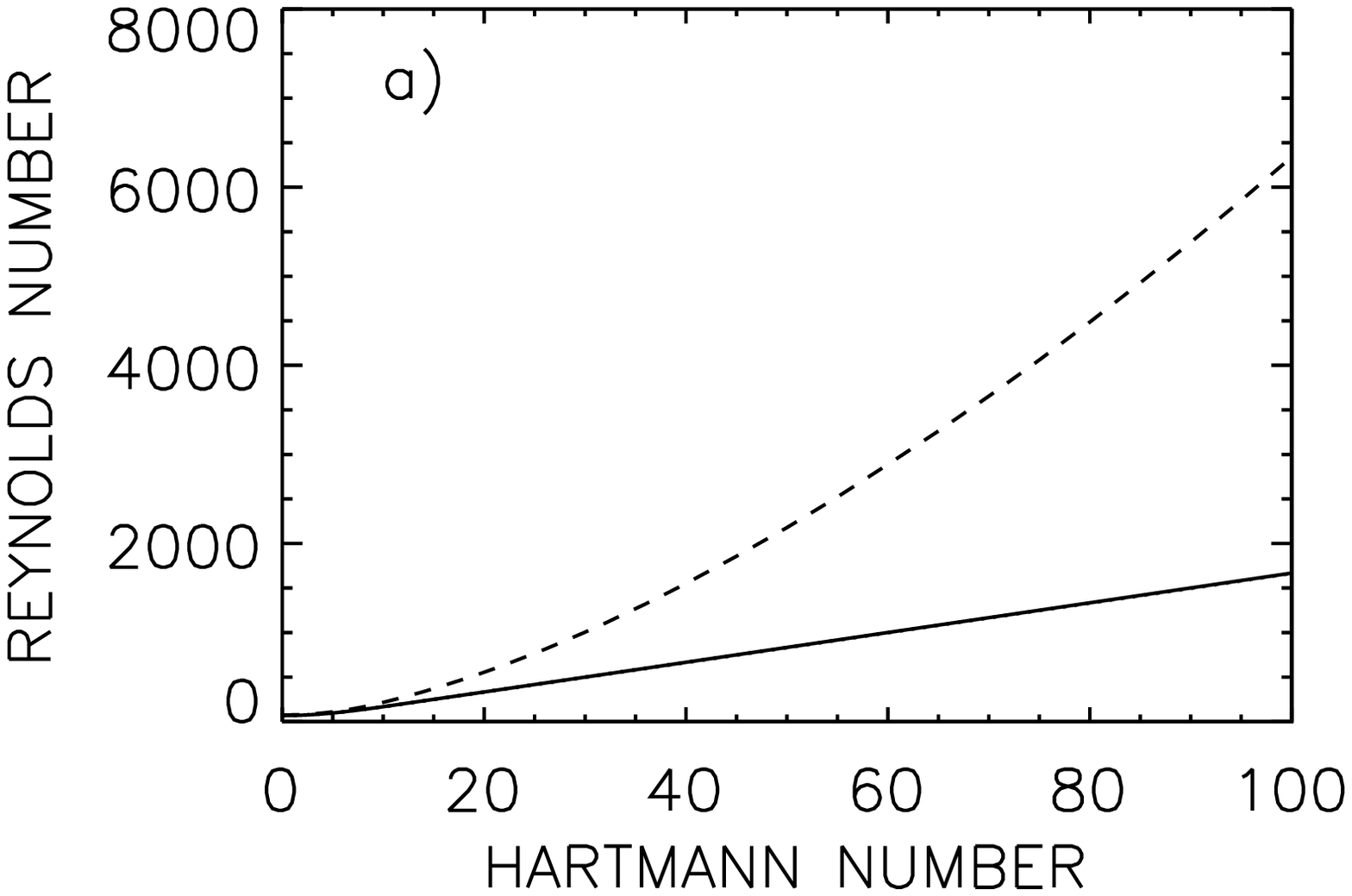,width=4.4cm,height=10cm}\hfill
\psfig{figure=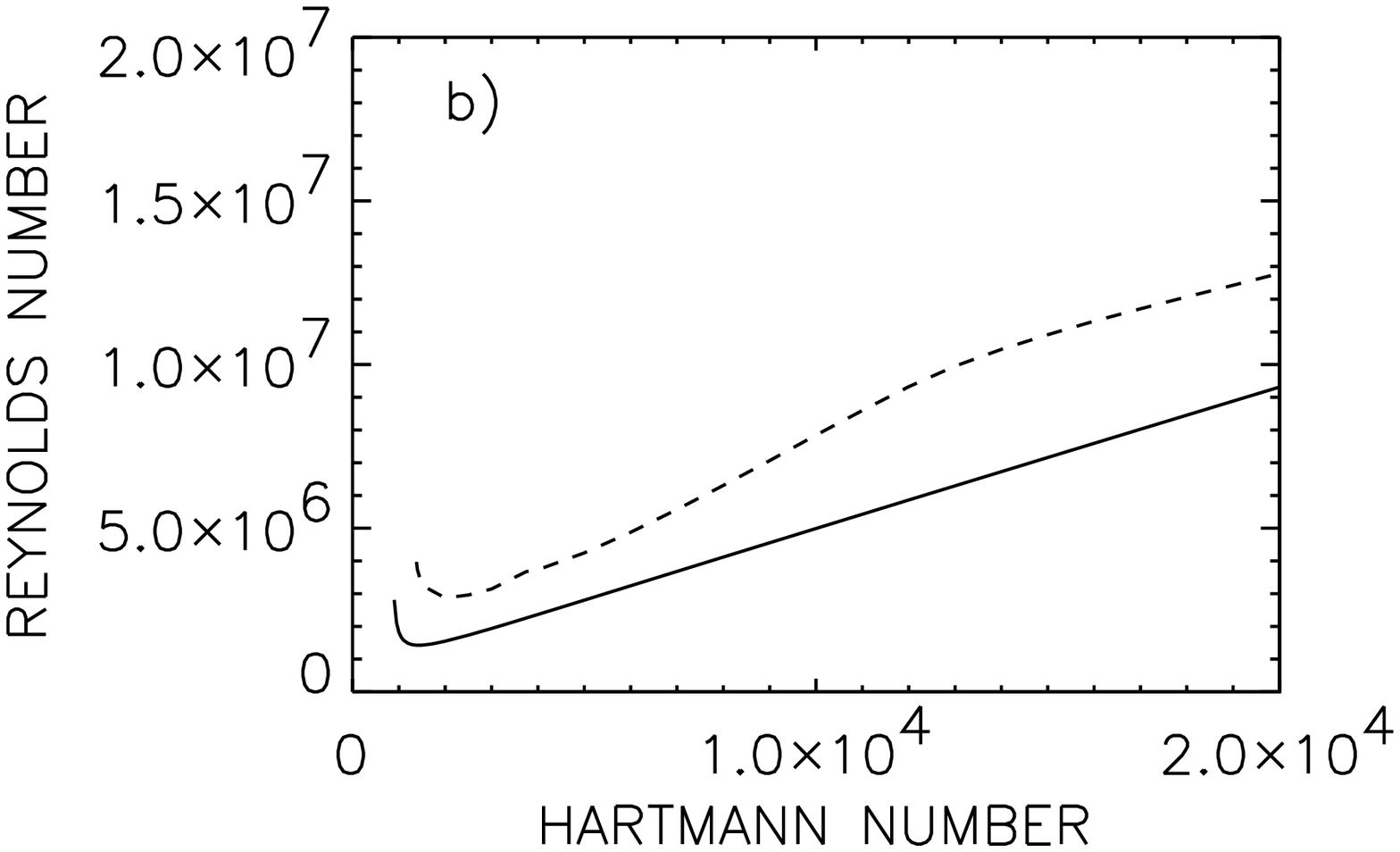,width=4.4cm,height=10cm}}
\caption{\label{f1} Insulating walls (vacuum): Stability lines for 
axisymmetric ($m=0$, solid lines)
and nonaxisymmetric instability modes with $m=1$  (dashed).  LEFT: resting outer cylinder, RIGHT: rotating outer
cylinder (33\%). }
\end{figure}
\begin{figure}[ht]
\hbox{
\psfig{figure=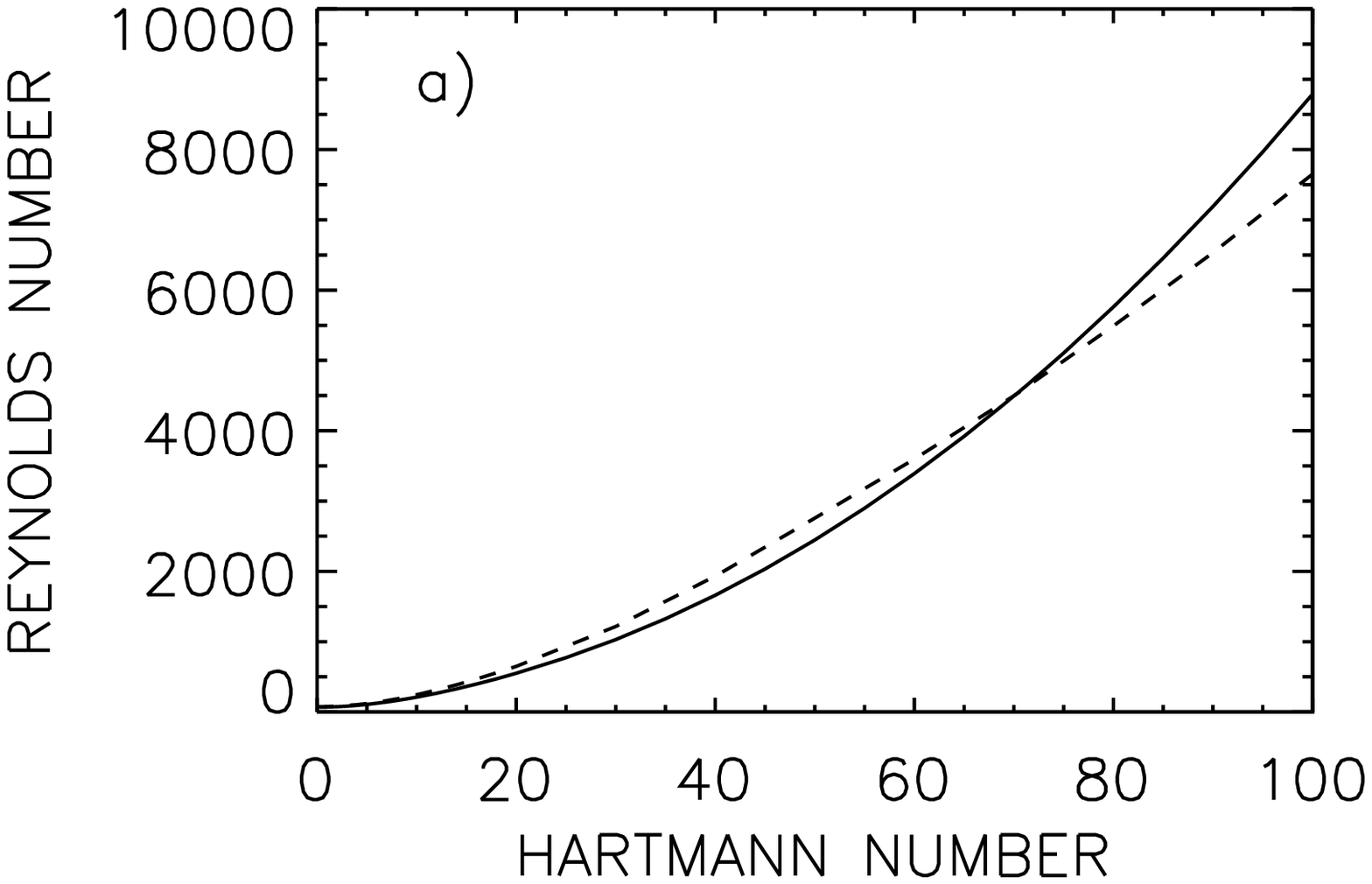,width=4.2cm,height=10cm}\hfill
\psfig{figure=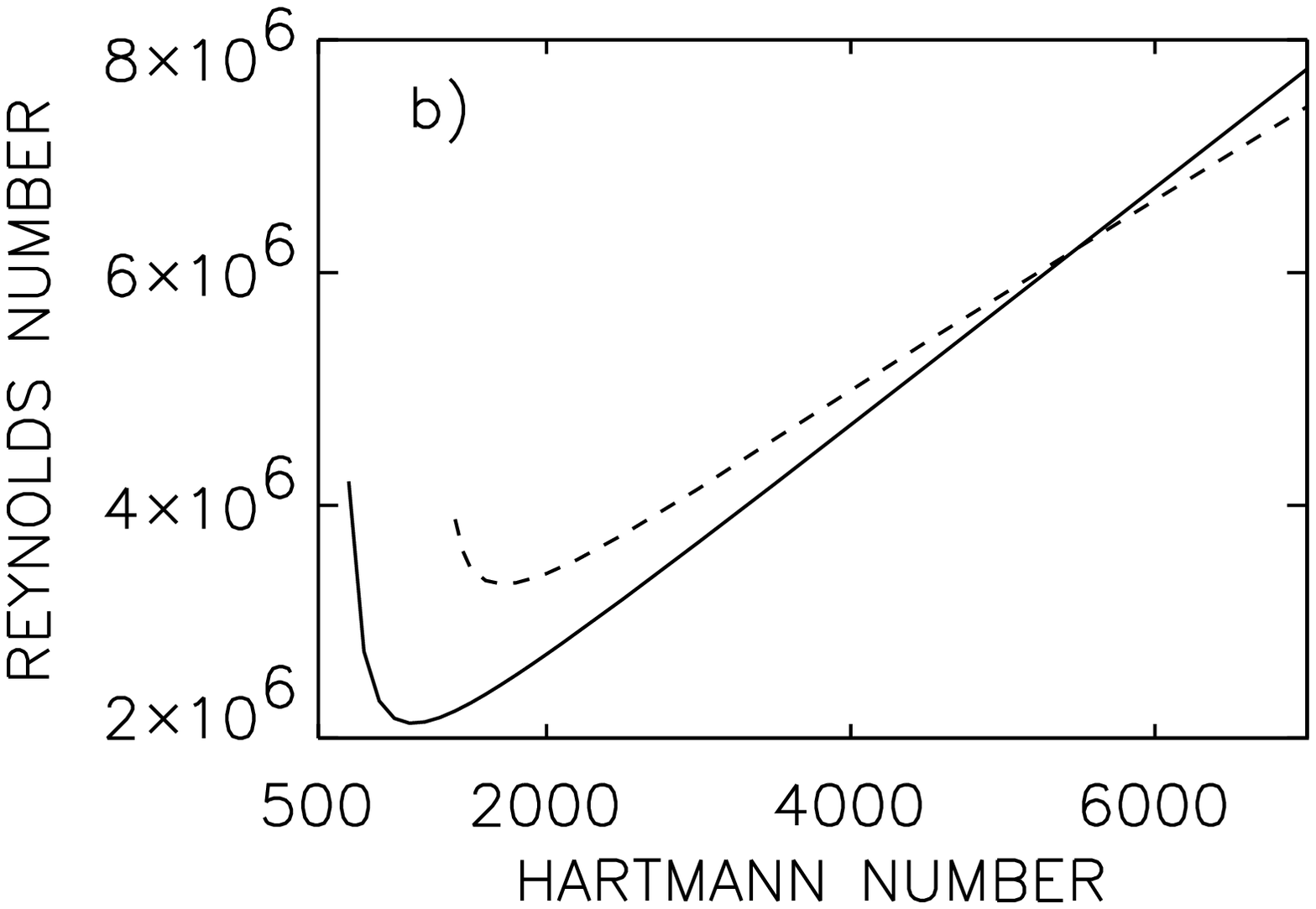,width=4.2cm,height=10cm}}
\caption{\label{f2} Conducting walls: Stability lines for 
axisymmetric ($m=0$, solid lines)
and nonaxisymmetric instability modes with $m=1$  (dashed).  LEFT: resting outer cylinder, RIGHT: rotating outer
cylinder (33 \%).}
\end{figure}
\begin{table}
\caption{\label{tab1} Coordinates of the absolute minima in 
Figs. \ref{f1} and \ref{f2} for rotating outer cylinder ($\hat\mu =0.33$) }
\medskip
\begin{tabular}{|l|c|c|}
\hline
 & conducting walls & insulating walls\\[0.5ex]
\hline
Reynolds  number & $2.13 \cdot 10^6$ & $1.42 \cdot 10^6$\\[1ex]
\hline
magnetic Reynolds  number & 21 & 14\\[1ex]
\hline
Hartmann number & $1100$ & $1400$\\[1ex]
\hline
Lundquist number & 3.47 & 4.42\\
\hline
\end{tabular}
\end{table}
Let us now concentrate to the small magnetic Prandtl number for liquid sodium,
i.e. Pm $=10^{-5}$.
We start with the results for  containers  with insulating walls and
outer cylinders at rest, (Fig. \ref{f1}a).  There are then linear instabilities even without magnetic fields.
For Ha$=0$ solutions for $m=0$ (Re$=68$) and $m=1$ (Re$=75$) are known, see
\cite{SRS02}. The axisymmetric mode possesses the
lowest eigenvalue. This is also true for nonvanishing magnetic field; we do
not find any crossover of the instability lines for axisymmetric and nonaxisymmetric modes. The
same is true for containers with rotating outer cylinder  (Fig. \ref{f1}b). For growing $\hat \mu$ the
Reynolds number for the hydrodynamic solution moves upwards, reaching
infinity for $\hat \mu = \hat\eta^2=0.25$ (here). The MRI is represented
by characteristic minima, in our case for $\hat\mu =0.33$ at Hartmann numbers of
order $10^3$ and Reynolds numbers of order $10^6$. The exact coordinates  of the
minima are given in Table \ref{tab1}.

In
order to characterize the Hartmann numbers note that for liquid sodium
\begin{equation}
B=2.2 {{\rm Ha} \over R_{\rm out}/ 10 {\rm cm}} \ {\rm Gauss}.
\label{B2.22}
\end{equation}
Hence, for $R_{\rm out} \simeq 22$ the magnetic field and the
Hartmann number have the same numerical values. 
With $\nu=10^{-2}$ cm$^2$/s and  $\hat \eta$=0.5 it follows from (\ref{RE})
and (\ref{HA})
\beg 
f_{\rm in}= 64{{\rm Re}/10^6\over \left(R_{\textrm out}/10{\textrm cm}\right)^2} \ {\textrm Hz}
\label{fff}
\ende
for the frequency of the inner cylinder. Hence, a container with  an outer
radius of 22 cm (see above) and an inner radius of 11 cm  filled with liquid
sodium and embedded in vacuum
requires a {\em rotation of about 19 Hz} in order to find the MRI. Following 
(\ref{B2.22}) 
the required magnetic field is about 1400 Gauss.

\begin{figure}[ht]
\psfig{figure=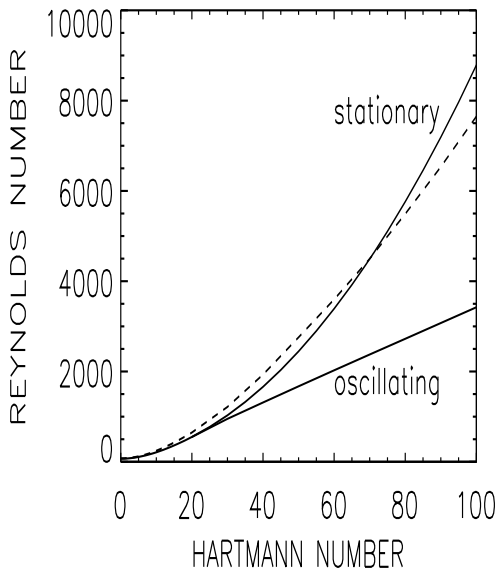,width=7cm,height=9cm}
\caption{\label{over}The same as in Fig. \ref{f2}a but with the inclusion 
of oscillating axisymmetric modes (overstability)  appearing here for lower Reynolds 
numbers.}
\end{figure}
The results for containers with {\em conducting walls} are given in Fig. 
\ref{f2}. Note that the minimal Reynolds numbers given in Fig. 
\ref{f2}b are
 higher than for insulating cylinder walls. The influence of the boundary conditions
 is not as small as expected. The main difference between the
two sorts of boundary conditions, however,  is the existence of crossovers of the instability lines for $m=0$ and $m=1$ in case of conducting walls. For both
resting and rotating outer cylinder Hartmann numbers exist above which the
nonaxisymmetric mode possesses a lower Reynolds number than the
axisymmetric mode. We have already shown the existence of such crossovers for conducting walls for
$1\leq {\rm Pm} \leq 0.01$ in \cite{SRS02}. It is now clear that the occurrence
of nonaxisymmetric solutions    as the preferred modes is a rather general phenomenon
for containers with conducting walls which can become important for the 
design of future dynamo experiments (Cowling theorem).
\subsection{Excitation of oscillating modes}
There are not only stationary patterns of flow and field possible but  the 
instability can also onset in form of oscillating solutions. This effect 
is called overstability. In case of rotating convection between two 
layers heated from below the onset of the instability in form of 
oscillating solutions  even possesses the 
lowest eigenvalues for certain Prandtl numbers, \cite{C61}.  We find a very similar behavior for the 
MHD Taylor-Couette flow between conducting cylinders for resting outer 
cylinder (see Fig. \ref{over}). It is a pair of waves traveling in
positive and negative $z$-direction. Note that the cylinder considered here
has no bound in vertical direction. If the cylinder is finite, however, the
possibility exists that traveling waves might be combined to standing
waves.


\end{document}